\documentclass[achemso,a4paper,twocolumn]{revtex4}
\usepackage{dcolumn}
\usepackage{amsmath}
\usepackage{graphicx}
\usepackage{latexsym}  
\usepackage{amsfonts}  
\usepackage{amssymb}  
\usepackage{mciteplus}
\DeclareGraphicsExtensions{.eps,.pdf,.gif,.jpg}  
  
\newcommand{\be}{\begin{equation}}
\newcommand{\ee}{\end{equation}}
\newcommand{\bea}{\begin{eqnarray}}
\newcommand{\eea}{\end{eqnarray}}
\newcommand{\Id}[1] {\int \! \! {\rm d}^3 #1}
\renewcommand{\vr} {{\bf r}}
\newcommand{\nn} {\nonumber}

\begin{document}

\title{Dependence of response functions and orbital functionals 
on occupation numbers} 
 
\author{S. Kurth$^{1,2}$}  
\author{C. R. Proetto$^{1,3,4}$}   
\author{K. Capelle$^{5,1}$}  
\affiliation{$^1$Institut f\"ur Theoretische Physik, Freie Universit\"at Berlin,   
Arnimallee 14, D-14195 Berlin, Germany}  
\affiliation{$^2$Nano-Bio Spectroscopy Group and 
European Theoretical Spectroscopy Facility (ETSF), 
Dpto. de F\'{i}sica de Materiales, 
Universidad del Pa\'{i}s Vasco UPV/EHU, Centro Mixto CSIC-UPV/EHU, 
Av. Tolosa 72, E-20018 San Sebasti\'{a}n, Spain} 
\affiliation{$^3$European Theoretical Spectroscopy Facility (ETSF)}
\affiliation{$^4$Centro At\'omico Bariloche and Instituto Balseiro, 
8400 S. C. de Bariloche, R\'{i}o Negro, Argentina}    
\affiliation{$^5$Departamento de F\'{\i}sica e Inform\'{a}tica, Instituto de 
F\'{\i}sica de S\~{a}o Carlos, Universidade de S\~{a}o Paulo, 
Caixa Postal 369, 13560-970 S\~{a}o Carlos, S\~{a}o Paulo, Brazil}

\date{\today}  

\begin{abstract}
Explicitly orbital-dependent approximations to the exchange-correlation energy 
functional of density functional theory typically not only depend on the 
single-particle Kohn-Sham orbitals but also on their occupation numbers in 
the ground state Slater determinant. The variational calculation of the 
corresponding exchange-correlation potentials with the Optimized Effective 
Potential (OEP) method therefore also requires a variation of the occupation 
numbers with respect to a variation in the effective single-particle 
potential, which is usually not taken into account. Here it is shown under 
which circumstances this procedure is justified. 
\end{abstract}
  
  
\maketitle  
  
\section{Introduction}

The central quantity of density functional theory 
\cite{HohenbergKohn:64,KohnSham:65}, the exchange-correlation energy $E_{xc}$, 
is a unique (though unknown) functional of the electron density. Popular 
approximations such as the local density approximation (LDA) and generalized 
gradient approximations (GGA's) express $E_{xc}$ as an {\em explicit} 
functional of the density. 

Recently, another class of approximations has attracted increasing interest:
{\em implicit} density functionals, expressing $E_{xc}$ as explicit functionals
of the Kohn-Sham single particle orbitals and energies and therefore only as 
implicit functionals of the density 
\cite{GraboKreibichKurthGross:00,KuemmelKronik:08}. Members of this class of 
functionals are the exact exchange functional (EXX), the popular hybrid 
functionals which mix GGA exchange with a fraction of exact exchange 
\cite{Becke:93,Becke:93-2,Becke:96,AdamoBarone:99}, the Perdew-Zunger
self-interaction correction \cite{PerdewZunger:81} and meta-GGA functionals 
\cite{PerdewKurthZupanBlaha:99,KurthPerdewBlaha:99,TaoPerdewStroverovScuseria:03} 
which include the orbital kinetic energy density as a key ingredient. 

At zero temperature, the orbital functionals mentioned above depend on the 
occupied orbitals only. Other functionals, such as the second-order correlation 
energy of G\"orling-Levy perturbation theory \cite{GoerlingLevy:93}, in 
addition depend explicitly on the unoccupied orbitals and the orbital 
energies. Moreover, all these orbital functionals are not only explicit 
functionals of the orbitals but also explicit functionals of the occupation 
numbers which, in turn, depend on the single-particle orbital energies. 
This additional energy dependence is ignored in common implementations of 
orbital- or energy-dependent functionals.

In order to calculate the single-particle Kohn-Sham potential corresponding to 
a given orbital functional, the so-called Optimized Effective Potential (OEP) 
method is used 
\cite{TalmanShadwick:76,GraboKreibichKurthGross:00,KuemmelKronik:08}. The 
OEP method is a variational method which aims to find that local potential 
whose orbitals minimize the given total energy expression. In principle, when 
performing the variation of the local potential one not only should vary the 
orbitals but also the orbital energies and occupation numbers. Typically, 
however, the variation with respect to the occupation numbers is not 
explicitly performed. In this work we will investigate when and why this is 
justified.

\section{Density Response Function}
\label{resp}

In this Section we analyze the problem of the eigenvalue 
dependence of the occupation numbers in the density and the non-interacting  
static linear density response function for various situations. We 
consider the case of zero temperature and distinguish between variations at 
fixed and variable particle number, i.e., for the canonical and grand-canonical 
ensemble. 

\subsection{Fixed particle number}
\label{fixed-N}

The density of $N$ non-interacting electrons (at zero temperature) moving in 
some electrostatic potential $v_s(\vr)$ is given by
\be
n(\vr) = \sum_{i}^{occ} | \varphi_i(\vr) |^2,
\label{dens}
\ee
where the single-particle orbitals are solutions of the Schr\"odinger equation 
\be
\left( - \frac{\nabla^2}{2} + v_s(\vr) \right) \varphi_i(\vr) = 
\varepsilon_i \varphi_i(\vr),
\label{schroedinger}
\ee
and the sum in Eq.~(\ref{dens}) runs over the $N$ occupied orbitals of the 
$N$-electron Slater determinant. For the 
ground state density one can rewrite Eq.~(\ref{dens}) as 
\be
n(\vr) = \sum_{i} \theta(\varepsilon_F - \varepsilon_i) | \varphi_i(\vr) |^2 
= \sum_{i} f_i | \varphi_i(\vr) |^2, 
\label{dens-2}
\ee
where the sum now runs over {\em all} orbitals. $\varepsilon_F$ is the Fermi 
energy, $\theta(x)$ is the Heavyside step function, and 
$f_i=\theta(\varepsilon_F - \varepsilon_i)$ is the occupation number of 
orbital $\varphi_i(\vr)$. It is evident from Eq.~(\ref{dens-2}) that the 
density not only depends on the (occupied) orbitals $\varphi_i(\vr)$ but also 
on the orbital energies $\varepsilon_i$, since the very specification of which
orbitals are occupied and which unoccupied depends on their energies. 

Through Eq.~(\ref{schroedinger}), 
both of these quantities are functionals of the potential $v_s(\vr)$, i.e., 
$\varphi_i(\vr) = \varphi_i[v_s](\vr)$, $\varepsilon_i = \varepsilon_i[v_s]$. 
The static density response function, which is the functional derivative of 
$n$ with respect to $v_s$, is therefore given as
\be
\tilde{\chi}(\vr,\vr') = \frac{\delta n(\vr)}{\delta v_s(\vr')} = 
\sum_i \frac{\delta f_i}{\delta v_s(\vr')} | \varphi_i(\vr) |^2 + 
\chi(\vr,\vr'),
\label{response-1}
\ee
with
\bea
\lefteqn{
\chi(\vr,\vr') = 
\sum_i f_i \left( \frac{\delta \varphi_i(\vr)}{\delta v_s(\vr')} 
\varphi_i^*(\vr) + c.c. \right) } \nn \\
&=& \sum_{\stackrel{i,k}{i \neq k}} f_i \left( \frac{\varphi_k^*(\vr) 
\varphi_k(\vr') \varphi_i(\vr) \varphi_i^*(\vr')}{\varepsilon_i - 
\varepsilon_k} + c.c. \right) \; .
\label{response-2}
\eea
The last step follows from first order perturbation theory, which can be 
used to obtain 
\be
\frac{\delta \varphi_i(\vr)}{\delta v_s(\vr')} = 
\sum_{\stackrel{k}{k \neq i}} \frac{\varphi_k(\vr) \varphi_k^*(\vr') 
\varphi_i(\vr')}{\varepsilon_i - \varepsilon_k}.
\label{dphi-dv}
\ee
For simplicity, we have assumed a non-degenerate single-particle spectrum. 

Usually, $\chi(\vr,\vr')$ of Eq.~(\ref{response-2}) is taken as the static 
density response function instead of $\tilde{\chi}$. Both expressions differ 
by the first term on the right hand side of Eq.~(\ref{response-1}), becoming
identical only if this term vanishes. In order to see when and how this happens
we consider two cases. 

{\em Case 1} comprises systems for which the single-particle spectrum has a 
finite gap between the highest occupied orbital (eigenvalue $\varepsilon_{N})$ 
and the lowest unoccupied orbital (eigenvalue $\varepsilon_{N+1})$. Then 
the Fermi energy $\varepsilon_F$ lies strictly between these two orbital 
energies, $\varepsilon_{N} < \varepsilon_F < \varepsilon_{N+1}$. Within 
the single-particle gap, the position of $\varepsilon_F$ is arbitrary (at zero 
temperature). The important point now is that upon (infinitesimal) variation 
of the potential $v_s$, $\epsilon_F$ remains fixed and does not need to be 
varied. The reason is that the variation $\delta \varepsilon_{N}$ of 
$\varepsilon_{N}$ due to the variation of $v_s$ is infinitesimal as well 
and $\varepsilon_F$ can be chosen such that 
$\varepsilon_F > \varepsilon_{N} + \delta \varepsilon_{N}$, 
thus leaving the particle number unchanged. Then the functional derivative of 
the occupation number with respect to $v_s$ becomes 
\be
\frac{\delta f_i}{\delta v_s(\vr)} = \frac{\partial \theta(\varepsilon_F - 
\varepsilon_i)}{\partial \varepsilon_i} \frac{\delta \varepsilon_i}{\delta 
v_s(\vr)} = - \delta(\varepsilon_F - \varepsilon_i) | \varphi_i(\vr)|^2,
\label{delta-occ}
\ee
where $\delta(x)$ is the Dirac delta function and we used the relation 
\be
\frac{\delta \varepsilon_i}{\delta v_s(\vr)} = |\varphi_i(\vr)|^2, 
\ee
which can be obtained from first-order perturbation theory. In the present 
case, the Fermi energy (which is in the single-particle gap) is not equal to 
any of the single-particle energies, the delta function in  
Eq.~(\ref{delta-occ}) vanishes and $\tilde{\chi}(\vr,\vr')$ of 
Eq.~(\ref{response-1}) coincides with the usual form of the static density 
response function of Eq.~(\ref{response-2}).

{\em Case 2} is the case of a vanishing single-particle gap, i.e., the 
case of an open-shell or metallic system. For notational simplicity, in 
the following discussion we still work with the assumption of a non-degenerate 
single-particle spectrum. Of course, particularly for open-shell systems, 
this assumption is inappropriate. The more general case including degenerate 
single-particle orbitals is discussed in Appendix \ref{append}.

The crucial difference to case 1 is 
that an infinitesimal variation of the potential $v_s$ now not only leads to a 
variation $\delta \varepsilon_i$ of the single-particle energies but also to 
a variation $\delta \varepsilon_F$ of the Fermi energy. This latter variation 
has to be taken into account in order for the particle number to be conserved 
(i.e., the infinitesimal variation $\delta N$ of the particle number upon 
variation of the potential strictly has to vanish, $\delta N = 0$). Then the 
functional derivative of the occupation number with respect to the potential 
consists of two terms and reads 
\bea
\frac{\delta f_i}{\delta v_s(\vr)} &=& \frac{\partial \theta(\varepsilon_F - 
\varepsilon_i)}{\partial \varepsilon_F} \frac{\delta \varepsilon_F}{\delta 
v_s(\vr)} + \frac{\partial \theta(\varepsilon_F - 
\varepsilon_i)}{\partial \varepsilon_i} \frac{\delta \varepsilon_i}{\delta 
v_s(\vr)}  \nn \\
&=& \delta(\varepsilon_F - \varepsilon_i) \left( | \varphi_F(\vr)|^2 
- | \varphi_i(\vr)|^2 \right),
\label{delta-occ-2}
\eea
where $\varphi_F$ is the highest occupied orbital with orbital energy equal to 
the Fermi energy. Due to the delta function, the r.h.s. of 
Eq.~(\ref{delta-occ-2}) vanishes and again $\tilde{\chi}(\vr,\vr')$ of 
Eq.~(\ref{response-1}) coincides with the static density response function 
$\chi(\vr,\vr')$ of the form given in Eq.~(\ref{response-2}).
 
From Eq.~(\ref{response-1}) the linear change in the density due to the 
perturbation $\delta v_s(\vr)$ is 
$\delta n(\vr) = \Id{r'} \tilde{\chi}(\vr,\vr')\delta {v_{s}}(\vr')$. One can 
then check explicitly that the result $\tilde{\chi}(\vr,\vr') = 
{\chi}(\vr,\vr') $ obtained here is fully consistent with a fixed number of 
particles: 
\bea
\delta N &=& \Id{r} ~ \delta n(\vr) =  \Id{r'} \delta {v_{s}}(\vr') \Id{r} ~ 
\tilde{\chi}(\vr,\vr')\nn \\ 
&=&  \Id{r'} \delta {v_{s}}(\vr') \Id{r} ~ \chi(\vr,\vr') = 0, 
\label{density-pert}
\eea
where the last equality follows from the orthonormality of the single-particle 
orbitals. 

\subsection{Grand canonical ensemble}

The analysis is slightly altered if the system of non-interacting electrons is 
connected to a particle bath, i.e., for the grand canonical ensemble 
characterized by a chemical potential $\mu$. The density (at zero temperature) 
is then given by
\be
n(\vr) = \sum_{i} \theta(\mu - \varepsilon_i) | \varphi_i(\vr) |^2 
= \sum_{i} f_i | \varphi_i(\vr) |^2, 
\label{dens-3}
\ee
where the occupation number now is given by $f_i = \theta(\mu - \varepsilon_i)$ 
and the sum again runs over all single-particle orbitals. When varying the 
occupation numbers with respect to variations of the potential, the chemical 
potential remains constant, independent of the single-particle spectrum having 
a finite or vanishing gap at $\mu$. The variation of $f_i$ then is obtained 
similarly to case 1 of the previous subsection as
 \be
\frac{\delta f_i}{\delta v_s(\vr)} = \frac{\partial \theta(\mu - 
\varepsilon_i)}{\partial \varepsilon_i} \frac{\delta \varepsilon_i}{\delta 
v_s(\vr)} = - \delta(\mu - \varepsilon_i) | \varphi_i(\vr)|^2 \; .
\label{delta-occ-3}
\ee
This term does not vanish if the chemical potential is aligned with one of the 
single-particle energies and the static density response function for the 
grand-canonical ensemble reads
\be
\tilde{\chi}(\vr,\vr') = \chi(\vr,\vr') - 
\sum_i \delta(\mu - \varepsilon_i) | \varphi_i(\vr) |^2 \; .
\label{response-3}
\ee

It is worth noting that now, due to the second term on the r.h.s. of 
Eq.~(\ref{response-3}), $\delta N$ (Eq.~(\ref{density-pert})) is 
different from zero which is of course consistent with the fact that here 
we are dealing with an open system.

\section{Implications for the Optimized Effective Potential}

The central idea of density functional theory is to write the ground state 
energy $E_{tot}$ of $N$ interacting electrons moving in an external 
electrostatic potential $v_0(\vr)$ as a functional of the ground-state 
density. This energy functional may then be split into various pieces as 
\be
E_{tot} = T_s[n] + \Id{r} \; v_0(\vr) n(\vr) + U[n] + E_{xc}[n],
\label{etot}
\ee 
where $T_s[n]$ is the kinetic energy functional of {\em non-interacting} 
electrons, 
\be
U[n] = \frac{1}{2} \Id{r} \Id{r'} \frac{n(\vr) n(\vr')}{|\vr -\vr'|}
\ee
is the classical electrostatic energy and $E_{xc}$ is the 
exchange-correlation energy functional which incorporates all complicated 
many-body effects and in practice has to be approximated. Minimization of 
Eq.~(\ref{etot}) with respect to the density leads to an effective 
single-particle equation of the form of Eq.~(\ref{schroedinger}) where the 
effective potential is
\be
v_s(\vr) = v_0(\vr) + \Id{r'} \frac{n(\vr')}{|\vr - \vr'|} + v_{xc}(\vr), 
\ee
with the exchange-correlation potential 
\be
v_{xc}(\vr) = \frac{\delta E_{xc}}{\delta n(\vr)} \; .
\ee

While the most popular approximations to the exchange-correlation energy 
$E_{xc}$ are explicit functionals of the density, 
there has been increasing interest in another class of approximations which 
are are only {\em implicit} functionals of the density. These functionals 
instead depend explicitly on the Kohn-Sham single-particle orbitals as well 
as on the Kohn-Sham orbital energies. One example for such a functional 
is the exact exchange energy given as
\be
E_x^{EXX} = - \frac{1}{4} \Id{r} \Id{r'} \frac{|\gamma(\vr,\vr')|^2}{|\vr - 
\vr'|},
\label{exx}
\ee
where
\be
\gamma(\vr,\vr') = \sum_i f_i \varphi_i(\vr) \varphi_i^*(\vr')
\ee
is the single-particle density matrix. As one can see, $E_x^{EXX}$ depends on 
the single-particle energies through the occupation numbers $f_i$. 
Other functionals such as, e.g., the correlation energy functional of 
second-order G\"orling-Levy perturbation theory \cite{GoerlingLevy:93}, 
depend on the orbital energies also in other ways (see below). 

In order to distinguish a genuine dependence on orbital energies from 
a dependence on occupation numbers we write for a general exchange-correlation 
energy functional $E_{xc} = E_{xc}[\{\varphi_i\},\{\varepsilon_i\},\{f_i\}]$. 
The exchange-correlation potential of such a functional can be computed by 
using the chain rule of functional differentiation as 
\be
v_{xc}(\vr) = \frac{\delta E_{xc}}{\delta n(\vr)} = \Id{r'} 
\frac{\delta E_{xc}}{\delta v_s(\vr')} \frac{\delta v_s(\vr')}{\delta n(\vr)}.  
\ee
Acting with the density response operator (\ref{response-1}) on both sides of 
this equation one arrives at
\bea
\lefteqn{ \Id{r'} v_{xc}(\vr') \tilde{\chi}(\vr',\vr) = \Id{r'} 
\frac{\delta E_{xc}}{\delta v_s(\vr')} } \nn\\
&=& \sum_i \Id{r'} \Bigg( \left( \frac{\delta E_{xc}}{\delta \varphi_i(\vr')} 
\bigg\vert_{\{\varepsilon_k\},\{f_k\}} 
\frac{\delta \varphi_i(\vr')}{\delta v_s(\vr)} + c.c. \right)\nn \\
&& +  \frac{\partial E_{xc}}{\partial \varepsilon_i} \bigg\vert_{\{\varphi_k\},
\{f_k\}} 
\frac{\delta \varepsilon_i}{\delta v_s(\vr)} \nn \\
&& +  
\frac{\partial E_{xc}}{\partial f_i} \bigg\vert_{\{\varphi_k\},\{\varepsilon_k\}} 
\frac{\delta f_i}{\delta v_s(\vr)} \Bigg) \; .
\label{oep-general}
\eea
In the last step we have used the chain rule once again and we also emphasize 
in the notation that when varying with respect to one set of variables 
(orbitals, orbital energies or occupation numbers) the other variables 
remain fixed. 

Eq.~(\ref{oep-general}) is the OEP integral equation in its general form. 
For a given approximate $E_{xc}$, this equation {\em defines} the 
corresponding $v_{xc} (\vr)$ and has to be solved in a self-consistent way 
together with the Kohn-Sham equations (Eq.~(\ref{schroedinger})). 
It differs in three ways from the form most commonly found in the literature 
(see, e.g., Refs.~\onlinecite{GraboKreibichKurthGross:00,KuemmelKronik:08} and 
references therein). One, the explicit energy dependence, is handled in a 
similar way as is the orbital dependence, via the chain rule. The other two 
arise from the implicit energy dependence of the occupation numbers, and 
are our main concern here. Similar to the discussion in the previous section 
we will again distinguish between the two cases of fixed particle number and 
systems in contact with a particle bath and discuss the role of these 
extra terms in both cases.

\subsection{Fixed particle number}

As we have seen in Section \ref{resp}, for the case of fixed par\-ticle number 
at zero temperature the functional derivative $\delta f_i / \delta v_s(\vr)$ 
vanishes both for systems with a finite and vanishing HOMO-LUMO gap. 
This has two consequences for Eq.~(\ref{oep-general}): 
first, we can replace the response function $\tilde{\chi}$ by the function 
$\chi$ of Eq.~(\ref{response-2}) and second, the last term on the r.h.s. 
of Eq.~(\ref{oep-general}) drops out. Therefore, the OEP equation reads
\bea
\lefteqn{ \Id{r'} v_{xc}(\vr') \chi(\vr',\vr) } \nn\\
&=& \sum_i \left[ \Id{r'} \left( \frac{\delta E_{xc}}{\delta \varphi_i(\vr')} 
\bigg\vert_{\{\varepsilon_k\},\{f_k\}} 
\frac{\delta \varphi_i(\vr')}{\delta v_s(\vr)} + c.c. \right) \right. \nn \\
&& + \left. \frac{\partial E_{xc}}{\partial \varepsilon_i} 
\bigg\vert_{\{\varphi_k\},\{f_k\}} 
\frac{\delta \varepsilon_i}{\delta v_s(\vr)}  \right] \; .
\label{oep-1}
\eea
This equation shows that despite the dependence of $E_{xc}$ on the 
occupation numbers (which, in turn, depend on the orbital energies), the 
variation with respect to these occupation numbers may be omitted for 
the calculation of the OEP integral equation for the exchange-correlation 
potential. This is, of course, what has been done in the vast majority of 
cases discussed in the literature. 

We note in passing that integrating Eq.~(\ref{oep-1}) over all space and 
using the orthornormality of the Kohn-Sham orbitals one can deduce the sum rule
\cite{EngelJiang:06}
\be
\sum_i \frac{\partial E_{xc}}{\partial \varepsilon_i} \bigg\vert_{\{\varphi_k\},
\{f_k\}} = 0 \; .
\label{sum-rule}
\ee

On quite general grounds, one expects that for an isolated system with
a fixed number of particles, $v_{xc} (\vr)$ is only defined up to a constant. 
To check if Eq.~(\ref{oep-1}) meets this condition we need an explicit
expression for $E_{xc}$. As a non-trivial example, we use 
\be
E_{xc} \approx E_x^{EXX} + {E_c}^{(2)}, 
\label{exc-gl2}
\ee
where $E_x^{EXX}$ is the exact exchange energy of Eq.~(\ref{exx}) and 
$E_c^{(2)}$ is the second-order correlation energy of G\"orling-Levy 
perturbation theory 
\cite{GoerlingLevy:93,EngelJiangFaccoBonetti:05,RigamontiProetto:06} 
defined by
\be
{E_c}^{(2)} = E_{c,1} + E_{c,2} \; ,
\label{GL2}
\ee
where
\be
E_{c,1} = \sum_{i,j} \frac {f_i(1-f_j)}{(\varepsilon_i-\varepsilon_j)}
\vert \langle i|v_x|j \rangle + \sum_k f_k (ik||kj)\vert^2 \; , 
\label{delta-HF}
\ee
and
\bea
E_{c,2} &=& \frac {1}{2} \sum_{i,j,k,l} 
\frac {f_if_j(1-f_k)(1-f_l)}{(\varepsilon_i + \varepsilon_j - 
\varepsilon_k - \varepsilon_l)} \nn \\
&& (ij||kl) \left[(kl||ij)-(kl||ji)\right].
\label{MP2}
\eea
In the equations above we have used the notations 
\be
(ij||kl) = \Id{r} \Id{r'} ~ \frac{{\varphi_i}^*(\vr) \varphi_k(\vr) 
{\varphi_j}^*(\vr') \varphi_l(\vr')}{|\vr - \vr'|} ,
\label{matrix-element}
\ee
and
\be
\langle i|v_x|j \rangle = \Id{r} ~ {\varphi_i}^*(\vr) v_x(\vr) \varphi_j(\vr)
\; .
\label{me-vx}
\ee
Suppose now that we introduce a rigid shift $v_s(\vr) \rightarrow v_s(\vr)+C$ 
in the effective single particle potential of Eq.~(\ref{schroedinger}). As a 
result, if $\{ \varphi_i \},\{ \varepsilon_i \},\{ f_i \}$ are a set 
of solutions for $v_s(\vr)$, the solutions for $v_s(\vr) + C$ are 
$\{ \varphi_i \},\{ \varepsilon_i + C \},\{ f_i \}$. This holds provided that 
Eq.~(\ref{oep-1}) determines $v_{xc}(\vr)$ only up to a constant. Inspection 
of Eq.~(\ref{oep-1}) confirms that this is the case: the l.h.s. is 
invariant under a rigid shift of $v_{xc}(\vr)$, and Eqs.~(\ref{exx}) and 
(\ref{GL2}) are invariant under the change
$\{ \varepsilon_i \} \rightarrow \{ \varepsilon_i + C \}$. 

\subsection{Grand canonical ensemble}

The situation is different if the system is in contact with a particle bath. 
Since in this case $\delta f_i/\delta v_s(\vr)$ does not vanish one has to 
use the full OEP equation (\ref{oep-general}). Here the dependence of both the 
density and the exchange-correlation energy on the occupation numbers has 
been taken into account explicitly when performing the variations and the 
two extra terms resulting from this variation cannot be neglected. 
Applications of this OEP formalism for open systems have been reported for 
quasi two-dimensional electron gases (2DEG) in $n$-doped semiconductor 
quantum wells where the $n$-doped regions act as particle reservoirs 
\cite{RigamontiReboredoProetto:03,RigamontiProettoReboredo:05,RigamontiProetto:07}. 

As another consequence of the extra terms, integration of 
Eq.~(\ref{oep-general}) over all space leads to the modified sum rule 
\bea
\lefteqn{
- \sum_i \delta(\mu - \varepsilon_i) \bar{v}_{xc,i} = 
\sum_i \bigg( \frac{\partial E_{xc}}{\partial \varepsilon_i} 
\bigg\vert_{\{\varphi_k\},\{f_k\}} } \nn \\
&& - \frac{\partial E_{xc}}{\partial f_i} \bigg\vert_{\{\varphi_k\},
\{\varepsilon_k\}} \delta(\mu - \varepsilon_i) \bigg) \; .
\label{sum-rule-mod}
\eea
where 
\be
\bar{v}_{xc,i} = \Id{r} \; v_{xc}(\vr) \; | \varphi_i(\vr)|^2 \; .
\label{vxcbar}
\ee
We take the exact exchange functional (\ref{exx}) as an example for a 
functional which does not explicitly depend on the single-particle energies. 
In this case, the first term on the r.h.s. of Eq.~(\ref{sum-rule-mod}) 
vanishes. If there exists a single-particle state whose energy equals 
the chemical potential, $\varepsilon_N = \mu$, we then obtain
\be
\bar{v}_{x,N}^{EXX} = \frac{\partial E_{x}^{EXX}}{\partial f_N} \; .
\ee
This relation is the complete analogue for the grand canonical ensemble 
of a well-known relation for fixed particle number which reads 
\cite{KriegerLiIafrate:92,LevyGoerling:96,KreibichKurthGraboGross:99}
\be
\bar{v}_{x,N}^{EXX} = \bar{u}_{x,N}^{EXX} \; ,
\ee
where
\be
\bar{u}_{x,N}^{EXX} = \frac{1}{f_N} \Id{r} \; \varphi_N(\vr) \; 
\frac{\delta E_x^{EXX}}{\delta \varphi_N(\vr)} \; .
\ee
For open 2DEG's, this relation has been obtained previously by studying
the asymptotic behavior of the exact-exchange potential 
\cite{RigamontiProettoReboredo:05}.

For the grand-canonical ensemble, $ v_{xc}(\vr)$ is 
{\em fully determined} by Eq.~(\ref{oep-general}) 
since this equation is {\em not} invariant under a rigid shift of the 
potential: the l.h.s. is not invariant due to the extra term in 
$ \tilde{\chi}(\vr,\vr') $ in Eq.~(\ref{response-3}). The r.h.s. is not 
invariant because $E_{xc}$ changes under the transformation 
$ \{ \varepsilon_i \} \rightarrow \{ \varepsilon_i + C \} $.  
This is due to the fact that the chemical potential $ \mu $ (which is 
determined by the particle reservoirs) remains fixed in the grand canonical 
ensemble and the above transformation leads to a change in the set of 
occupation numbers and self-consistent KS orbitals, $\{ f_i \}$ and 
$ \{ \varphi_i \}$, respectively.

\section{Conclusions}
In this work we have addressed the question why and when one can ignore 
the explicit dependence on the orbital occupation numbers (which in turn 
depend explicitly on the orbital energies) when calculating 
both the static linear density response function and the effective 
single-particle potential corresponding to an orbital-dependent 
exchange-correlation energy functional. We have shown that the variation of 
the occupation numbers may safely be neglected for systems with fixed 
particle number. For systems connected to a particle bath, however, this 
variation leads to non-vanishing contributions and needs to be taken into 
account. 

\appendix

\section{Degenerate single-particle spectrum}
\label{append}

In general, the single-particle spectrum will have eigenvalues which may be 
degenerate. In particular, in the case of open-shell systems, the energy of the 
highest occupied orbital is degenerate and the arguments of the {\em Case 2 } 
discussed in Section \ref{fixed-N} need to be modified. 

As degeneracy is in almost all cases related to symmetry we will use the 
language of group theory. In particular, the single-particle orbitals will be 
labelled by the a complete set of quantum numbers $\{n,l,m\}$ where $n$ is 
the principal quantum number (which is not related to symmetry), $l$ is a 
label denoting the irreducible representation of the symmetry group $\cal{G}$ 
of the single-particle potential $v_s(\vr)$, and $m$ labels a partner within 
that representation. The single-particle equation now reads
\be
\left( - \frac{\nabla^2}{2} + v_s(\vr) \right) \varphi_{nlm}(\vr) = 
\varepsilon_{nl} \varphi_{nlm}(\vr)
\label{schroed-2}
\ee
and it should be noted that the eigenvalue $\varepsilon_{nl}$ is independent 
of the partner label $m$. Furthermore, writing the energy eigenvalue as a 
functional of the potential, $\varepsilon_{nl}[v_s]$, one has to keep in mind 
that this functional is only well defined for potentials which are invariant 
under the transformations of the symmetry group $\cal{G}$ because $l$ refers 
to an irreducible representation of that group. Therefore, we calculate the 
variation of the orbital energies, $\delta \varepsilon_{nl} = 
\varepsilon_{nl}[v_s+ \delta v_s] - \varepsilon_{nl}[v_s]$ resulting from a 
variation $\delta v_s(\vr)$ which preserves the symmetry of 
$v_s(\vr)$. Replacing $v_s \to v_s + \delta v_s$, $\varphi_{nlm} \to 
\varphi_{nlm} + \delta \varphi_{nlm}$, and $\varepsilon_{nl} \to 
\varepsilon_{nl} + \delta \varepsilon_{nl} $ in Eq.~(\ref{schroed-2}), one 
finds that the first-order change in the energy eigenvalue is given by 
\be
\delta \varepsilon_{nl} = \Id{r} | \varphi_{nlm}(\vr)|^2 \delta v_s(\vr) \; .
\label{dele}
\ee
Summing this equation over the partner label $m$ one obtains
\be
d_{nl} \delta \varepsilon_{nl} = \Id{r}\sum_m | \varphi_{nlm}(\vr)|^2 
\delta v_s(\vr) \; ,
\ee
where $d_{nl}$ is the degeneracy of $\varepsilon_{nl}$. Now we note that the 
single-particle orbitals $\varphi_{nlm}(\vr)$ may be written as 
\be
\varphi_{nlm}(\vr) = R_{nl}(\vr) X_{lm}(\vr) \, ,
\label{phi-prod}
\ee
where $R_{nl}(\vr)$ is a totally symmetric function  which is invariant under 
all symmetry transformations $T$ of the group $\cal{G}$ and $X_{lm}(\vr)$ is a 
function which transforms according to the irreducible representation $l$ of 
$\cal{G}$, i.e., 
\be
X_{lm}(R^{-1}(T) \vr) = \sum_{m'} \Gamma^{(l)}(T)_{m'm} X_{lm'}(\vr) \; .
\ee
Here, $R(T)$ is a $3 \times 3$ matrix describing the symmetry operation 
$T \in \cal{G}$ in three dimensional space and $\Gamma^{(l)}(T)$ is the 
representation matrix of group element $T$ in the irreducible representation 
$l$ of $\cal{G}$. Noting now that $\sum_m |X_{lm}(\vr)|^2$ is a totally 
symmetric function, we find for the functional derivative
\be
\frac{\delta \varepsilon_{nl}}{\delta v_s(\vr)} = 
| \tilde{R}_{nl}(\vr)|^2 \; ,
\label{dele-delv}
\ee
where we have defined 
\be
\tilde{R}_{nl}(\vr) = \frac{1}{\sqrt{d_{nl}}} 
R_{nl}(\vr) \sqrt{\sum_m |X_{lm}(\vr)|^2} 
\ee
which is again invariant under all symmetry transformations of the group 
$\cal{G}$. 

Eq.~(\ref{dele-delv}) will shortly be used to repeat the arguments of Section 
\ref{fixed-N} for the degenerate, open-shell case. Before we do so, we  
point out that the definition of the density of Eq.~(\ref{dens-2}) needs to be 
modified because not all orbitals with energy $\varepsilon_F$ are (fully) 
occupied. This can be achieved, e.g., by writing the density as 
\be
n(\vr) = \sum_{n,l,m} f_{nlm}|\varphi_{nlm}(\vr)|^2
\ee
and occupying all degenerate orbitals of the partially 
filled subshell with the same fractional number of electrons, i.e., 
by defining the occupation number of the partially filled subshell by 
$f_{nlm}=f_{nl}= (n_{nl}/d_{nl}) \theta(\varepsilon_F - \varepsilon_{nl})$ where 
$n_{nl}$ is the number of electrons in the open subshell. 
With this definition, the static density response function reads 
\be
\tilde{\chi}(\vr,\vr') = \chi(\vr,\vr') + \sum_{n,l,m} 
\frac{\delta f_{nl}}{\delta v_s(\vr)} |\varphi_{nlm}(\vr)|^2
\ee
with
\bea
\lefteqn{
\chi(\vr,\vr') = \sum_{\stackrel{\stackrel{n,l,m}{n',l',m'}}{\varepsilon_{nl} \neq 
\varepsilon_{n'l'}}} f_{nl} } \nn \\
&& \!\!\!\!\!\!\!\!\left( \frac{\varphi_{n'l'm'}^*(\vr) \varphi_{n'l'm'}(\vr') 
\varphi_{nlm}(\vr) \varphi_{nlm}^*(\vr')}{\varepsilon_{nl} - 
\varepsilon_{n'l'}} + c.c. \right) \; .
\eea
and 
\bea
\lefteqn{
\frac{\delta f_{nl}}{\delta v_s(\vr)} = \frac{n_{nl}}{d_{nl}} 
\delta(\varepsilon_F - \varepsilon_{nl}) }\nn\\
&& \left( |\tilde{R}_F(\vr)|^2 - |\tilde{R}_{nl}(\vr)|^2 \right) = 0 \; , 
\eea
where the last equality follows because the total symmetric part of 
degenerate orbitals is identical. Therefore, just as in the non-degenerate 
case at fixed particle number, the functional derivative w.r.t. the occupation 
numbers may be neglected both in the calculation of the density response 
function as well as in the derivation of the OEP equation. 

\acknowledgments
S.K. acknowledges support through the Ikerbasque foundation. C.R.P. was
supported by the European Community through a Marie Curie Incoming 
International Fellowship (MIF1-CT-2006-040222) and CONICET of Argentina 
through PIP 5254. K.C. was supported by FAPESP and CNPq. We thank Stefano 
Pittalis, Hardy Gross, Angel Rubio, and Santiago Rigamonti for useful 
discussions.

\ifx\mcitethebibliography\mciteundefinedmacro
\PackageError{achemso.bst}{mciteplus.sty has not been loaded}
{This bibstyle requires the use of the mciteplus package.}\fi


\begin{mcitethebibliography}{23}
\providecommand*{\natexlab}[1]{#1}
\mciteSetBstSublistMode{f}
\mciteSetBstMaxWidthForm{subitem}{(\alph{mcitesubitemcount})}
\mciteSetBstSublistLabelBeginEnd{\mcitemaxwidthsubitemform\space}
{\relax}{\relax}

\bibitem[Hohenberg and Kohn(1964)]{HohenbergKohn:64}
Hohenberg,~P.; Kohn,~W. \emph{Phys.~Rev.} \textbf{1964}, \emph{136}, B864\relax
\mciteBstWouldAddEndPuncttrue
\mciteSetBstMidEndSepPunct{\mcitedefaultmidpunct}
{\mcitedefaultendpunct}{\mcitedefaultseppunct}\relax
\EndOfBibitem
\bibitem[Kohn and {L.J.~Sham}(1965)]{KohnSham:65}
Kohn,~W.; Sham,~L.J. \emph{Phys.~Rev.} \textbf{1965}, \emph{140}, A1133\relax
\mciteBstWouldAddEndPuncttrue
\mciteSetBstMidEndSepPunct{\mcitedefaultmidpunct}
{\mcitedefaultendpunct}{\mcitedefaultseppunct}\relax
\EndOfBibitem
\bibitem[Grabo et~al.(2000)Grabo, Kreibich, Kurth, and {E.K.U.
  Gross}]{GraboKreibichKurthGross:00}
Grabo,~T.; Kreibich,~T.; Kurth,~S.; Gross,~E.K.U., Orbital functionals in
  density functional theory: the optimized effective potential method.
  \emph{{S}trong {C}oulomb {C}orre\-lations in {E}lectronic {S}tructure
  {C}alculations: {B}eyond {L}ocal {D}ensity {A}pproxi\-mations}, Amsterdam,
  2000;
\newblock p 203\relax
\mciteBstWouldAddEndPuncttrue
\mciteSetBstMidEndSepPunct{\mcitedefaultmidpunct}
{\mcitedefaultendpunct}{\mcitedefaultseppunct}\relax
\EndOfBibitem
\bibitem[K{\"u}mmel and Kronik(2008)]{KuemmelKronik:08}
K{\"u}mmel,~S.; Kronik,~L. \emph{Rev.~Mod.~Phys.} \textbf{2008}, \emph{80},
  3\relax
\mciteBstWouldAddEndPuncttrue
\mciteSetBstMidEndSepPunct{\mcitedefaultmidpunct}
{\mcitedefaultendpunct}{\mcitedefaultseppunct}\relax
\EndOfBibitem
\bibitem[{A.D.~Becke}(1993)]{Becke:93}
Becke~A.D. \emph{J.~Chem. Phys.} \textbf{1993}, \emph{98}, 1372\relax
\mciteBstWouldAddEndPuncttrue
\mciteSetBstMidEndSepPunct{\mcitedefaultmidpunct}
{\mcitedefaultendpunct}{\mcitedefaultseppunct}\relax
\EndOfBibitem
\bibitem[{A.D.~Becke}(1993)]{Becke:93-2}
Becke~A.D. \emph{J.~Chem. Phys.} \textbf{1993}, \emph{98}, 5648\relax
\mciteBstWouldAddEndPuncttrue
\mciteSetBstMidEndSepPunct{\mcitedefaultmidpunct}
{\mcitedefaultendpunct}{\mcitedefaultseppunct}\relax
\EndOfBibitem
\bibitem[{A.D.~Becke}(1996)]{Becke:96}
Becke~A.D. \emph{J.~Chem. Phys.} \textbf{1996}, \emph{104}, 1040\relax
\mciteBstWouldAddEndPuncttrue
\mciteSetBstMidEndSepPunct{\mcitedefaultmidpunct}
{\mcitedefaultendpunct}{\mcitedefaultseppunct}\relax
\EndOfBibitem
\bibitem[Adamo and Barone(1999)]{AdamoBarone:99}
Adamo,~C.; Barone,~V. \emph{J.~Chem. Phys.} \textbf{1999}, \emph{110},
  6158\relax
\mciteBstWouldAddEndPuncttrue
\mciteSetBstMidEndSepPunct{\mcitedefaultmidpunct}
{\mcitedefaultendpunct}{\mcitedefaultseppunct}\relax
\EndOfBibitem
\bibitem[{J.P.~Perdew} and Zunger(1981)]{PerdewZunger:81}
Perdew,~J.P.; Zunger,~A. \emph{Phys.~Rev.~B} \textbf{1981}, \emph{23},
  5048\relax
\mciteBstWouldAddEndPuncttrue
\mciteSetBstMidEndSepPunct{\mcitedefaultmidpunct}
{\mcitedefaultendpunct}{\mcitedefaultseppunct}\relax
\EndOfBibitem
\bibitem[{J.P.~Perdew} et~al.(1999){J.P.~Perdew}, Kurth, Zupan, and
  Blaha]{PerdewKurthZupanBlaha:99}
Perdew,~J.P. ; Kurth,~S.; Zupan,~A.; Blaha,~P. \emph{Phys.~Rev. Lett.}
  \textbf{1999}, \emph{82}, 2544, ibid. {\bf 82}, 5179 (1999)(E)\relax
\mciteBstWouldAddEndPuncttrue
\mciteSetBstMidEndSepPunct{\mcitedefaultmidpunct}
{\mcitedefaultendpunct}{\mcitedefaultseppunct}\relax
\EndOfBibitem
\bibitem[Kurth et~al.(1999)Kurth, {J.P.~Perdew}, and
  Blaha]{KurthPerdewBlaha:99}
Kurth,~S.; Perdew,~J.P.; Blaha,~P. \emph{Int.~J.~Quantum Chem.}
  \textbf{1999}, \emph{75}, 889\relax
\mciteBstWouldAddEndPuncttrue
\mciteSetBstMidEndSepPunct{\mcitedefaultmidpunct}
{\mcitedefaultendpunct}{\mcitedefaultseppunct}\relax
\EndOfBibitem
\bibitem[Tao et~al.(2003)Tao, {J.P.~Perdew}, {V.N.~Staroverov}, and
  {G.E.~Scuseria}]{TaoPerdewStroverovScuseria:03}
Tao,~J.; Perdew,~J.P.; Staroverov,~V.N.; Scuseria,~G.E. \emph{Phys.~Rev.
  Lett.} \textbf{2003}, \emph{91}, 146401\relax
\mciteBstWouldAddEndPuncttrue
\mciteSetBstMidEndSepPunct{\mcitedefaultmidpunct}
{\mcitedefaultendpunct}{\mcitedefaultseppunct}\relax
\EndOfBibitem
\bibitem[G{\"o}rling and Levy(1993)]{GoerlingLevy:93}
G{\"o}rling,~A.; Levy,~M. \emph{Phys.~Rev.~B} \textbf{1993}, \emph{47},
  13105\relax
\mciteBstWouldAddEndPuncttrue
\mciteSetBstMidEndSepPunct{\mcitedefaultmidpunct}
{\mcitedefaultendpunct}{\mcitedefaultseppunct}\relax
\EndOfBibitem
\bibitem[{J.D.~Talman} and {W.F.~Shadwick}(1976)]{TalmanShadwick:76}
Talman,~J.D.; Shadwick,~W.F. \emph{Phys.~Rev.~A} \textbf{1976}, \emph{14},
  36\relax
\mciteBstWouldAddEndPuncttrue
\mciteSetBstMidEndSepPunct{\mcitedefaultmidpunct}
{\mcitedefaultendpunct}{\mcitedefaultseppunct}\relax
\EndOfBibitem
\bibitem[Engel and Jiang(2006)]{EngelJiang:06}
Engel,~E.; Jiang,~H. \emph{Int.~J.~Quantum Chem.} \textbf{2006}, \emph{106},
  3242\relax
\mciteBstWouldAddEndPuncttrue
\mciteSetBstMidEndSepPunct{\mcitedefaultmidpunct}
{\mcitedefaultendpunct}{\mcitedefaultseppunct}\relax
\EndOfBibitem
\bibitem[Engel et~al.(2005)Engel, Jiang, and
  {A.~Facco~Bonetti}]{EngelJiangFaccoBonetti:05}
Engel,~E.; Jiang,~H.; Facco~Bonetti,~A. \emph{Phys.~Rev.~A} \textbf{2005},
  \emph{72}, 052503\relax
\mciteBstWouldAddEndPuncttrue
\mciteSetBstMidEndSepPunct{\mcitedefaultmidpunct}
{\mcitedefaultendpunct}{\mcitedefaultseppunct}\relax
\EndOfBibitem
\bibitem[Rigamonti and {C.R.~Proetto}(2006)]{RigamontiProetto:06}
Rigamonti,~S.; Proetto,~C.R. \emph{Phys.~Rev.~B} \textbf{2006}, \emph{73},
  235319\relax
\mciteBstWouldAddEndPuncttrue
\mciteSetBstMidEndSepPunct{\mcitedefaultmidpunct}
{\mcitedefaultendpunct}{\mcitedefaultseppunct}\relax
\EndOfBibitem
\bibitem[Rigamonti et~al.(2003)Rigamonti, {F.A.~Reboredo}, and
  {C.R.~Proetto}]{RigamontiReboredoProetto:03}
Rigamonti,~S.; Reboredo,~F.A.; Proetto,~C.R. \emph{Phys.~Rev.~B}
  \textbf{2003}, \emph{68}, 235309\relax
\mciteBstWouldAddEndPuncttrue
\mciteSetBstMidEndSepPunct{\mcitedefaultmidpunct}
{\mcitedefaultendpunct}{\mcitedefaultseppunct}\relax
\EndOfBibitem
\bibitem[Rigamonti et~al.(2005)Rigamonti, {C.R.~Proetto}, and
  {F.A.~Reboredo}]{RigamontiProettoReboredo:05}
Rigamonti,~S.; Proetto,~C.R.; Reboredo,~F.A. \emph{Europhys.~Lett.}
  \textbf{2005}, \emph{70}, 116\relax
\mciteBstWouldAddEndPuncttrue
\mciteSetBstMidEndSepPunct{\mcitedefaultmidpunct}
{\mcitedefaultendpunct}{\mcitedefaultseppunct}\relax
\EndOfBibitem
\bibitem[Rigamonti and {C.R.~Proetto}(2007)]{RigamontiProetto:07}
Rigamonti,~S.; Proetto,~C.R. \emph{Phys.~Rev. Lett.} \textbf{2007},
  \emph{98}, 066806\relax
\mciteBstWouldAddEndPuncttrue
\mciteSetBstMidEndSepPunct{\mcitedefaultmidpunct}
{\mcitedefaultendpunct}{\mcitedefaultseppunct}\relax
\EndOfBibitem
\bibitem[{J.B.~Krieger} et~al.(1992){J.B.~Krieger}, Li, and
  {G.J.~Iafrate}]{KriegerLiIafrate:92}
Krieger,~J.B.; Li,~Y.; Iafrate,~G.J. \emph{Phys.~Rev.~A} \textbf{1992},
  \emph{45}, 101\relax
\mciteBstWouldAddEndPuncttrue
\mciteSetBstMidEndSepPunct{\mcitedefaultmidpunct}
{\mcitedefaultendpunct}{\mcitedefaultseppunct}\relax
\EndOfBibitem
\bibitem[Levy and G{\"o}rling(1996)]{LevyGoerling:96}
Levy,~M.; G{\"o}rling,~A. \emph{Phys.~Rev.~A} \textbf{1996}, \emph{53},
  3140\relax
\mciteBstWouldAddEndPuncttrue
\mciteSetBstMidEndSepPunct{\mcitedefaultmidpunct}
{\mcitedefaultendpunct}{\mcitedefaultseppunct}\relax
\EndOfBibitem
\bibitem[Kreibich et~al.(1999)Kreibich, Kurth, Grabo, and
  {E.K.U.~Gross}]{KreibichKurthGraboGross:99}
Kreibich,~T.; Kurth,~S.; Grabo,~T.; Gross,~E.K.U. \emph{Adv.~Quantum Chem.}
  \textbf{1999}, \emph{33}, 31\relax
\mciteBstWouldAddEndPuncttrue
\mciteSetBstMidEndSepPunct{\mcitedefaultmidpunct}
{\mcitedefaultendpunct}{\mcitedefaultseppunct}\relax
\EndOfBibitem
\end{mcitethebibliography}
\end{document}